\documentstyle[aps, prl, floats, epsfig,amsmath,amssymb ]{revtex}

\title{Approximate solution of variational wave functions for strongly
  correlated systems: Description of a correlated insulator }

\author{
  Bal\'azs Het\'enyi } \address{Institute for Theoretical Physics, Graz
  University of Technology, Petersgasse 16, A-8010 Graz, Austria
  \\ and \\
  Research Institute for Solid State Physics and Optics, Hungarian Academy of
  Sciences, 1525 Budapest 114, Pf 49, Hungary}

\begin{document}

\maketitle


\begin{abstract}
  An approximate solution scheme, similar to the Gutzwiller approximation, is
  presented for the Baeriswyl and the Baeriswyl-Gutzwiller variational
  wavefunctions.  The phase diagram of the one-dimensional Hubbard model as a
  function of interaction strength and particle density is determined.  For
  the Baeriswyl wavefunction a metal-insulator transition is found at
  half-filling, where the metallic phase ($U<U_c$) corresponds to the
  Hartree-Fock solution, the insulating phase is one with finite double
  occupations arising from bound excitons.  This transition can be viewed as
  the "inverse" of the Brinkman-Rice transition.  Close to but away from half
  filling, the $U>U_c$ phase displays a finite Fermi step, as well as double
  occupations originating from bound excitons.  As the filling is changed
  away from half-filling bound excitons are supressed.  For the
  Baeriswyl-Gutzwiller wavefunction at half-filling a metal-insulator
  transition between the correlated metallic and excitonic insulating state
  is found.  Away from half-filling bound excitons are suppressed quicker
  than for the Baeriswyl wavefunction.
\end{abstract}


\section{Introduction}

\label{sec:intro}

Variational studies have contributed greatly to our understanding of strongly
correlated systems, described by the Hubbard
model~\cite{Gutzwiller63,Gutzwiller65,Hubbard63,Kanamori63} and its
extensions.  While the last decades saw the development of the dynamical
mean-field theory~\cite{Georges96} and the density matrix renormalization
group~\cite{Schollwoeck05} variational studies still play an important role
in understanding metal-insulator transitions.  In part this is due to their
relative simplicity and applicability to large systems irrespective of number
of dimensions.  Two frequently used variational wavefunctions, are the
Gutzwiller wavefunction~\cite{Gutzwiller63,Gutzwiller65} (GWF) and the
Baeriswyl wavefunction~\cite{Baeriswyl86,Baeriswyl00} (BWF).  The former is
based on suppressing charge fluctuations in the noninteracting solution, the
latter on projecting out fluctuations in the hoppings from a completely
projected GWF.  The combined use of both projectors (Baeriswyl-Gutzwiller
wave function (BGWF) or Gutzwiller-Baeriswyl wave function
(GBWF))~\cite{Otsuka92,Dzierzawa95} has recently raised the possibility of
superconductivity in the two-dimensional Hubbard
model~\cite{Baeriswyl09,Eichenberger09}.  The idea of using projections based
on the kinetic energy or more general operators also appears in continuous
models~\cite{Vitiello88,Hetenyi99,Sarsa00}.

The GWF can be solved exactly only in one~\cite{Metzner87} and in
infinite~\cite{Metzner89} dimensions.  In one dimension the exact solution of
GWF is metallic, in contradiction with the exact result for the Hubbard
model~\cite{Lieb68}.  The GWF, however was shown to be metallic for all
finite dimensions~\cite{Millis91,Dzierzawa97}.  In higher finite dimensions
high order diagrammatic treatments~\cite{Gulacsi93} and quantum Monte
Carlo~\cite{Yokoyama87} are possible.  Only extended versions of the GWF can
account for insulating behavior: when correlations between doubly occupied
sites and empty sites are incorporated (bound
excitons)~\cite{Baeriswyl00,Capello05,Tahara08}, or when the non-interacting
wavefunction from which charge fluctuations are projected out is itself
insulating~\cite{Kollar01,Kollar02}.

For the BWF an exact analytical solution is in general not available.  It can
be shown~\cite{Dzierzawa97} that the Drude weight is identically zero, hence
the BWF is an insulating wavefunction.  If the N\'{e}el state is assumed to
be the wave function for infinite interaction then a solution is
feasible.~\cite{Dzierzawa97} For the general case analytical approximations
exist~\cite{Baeriswyl86,Dzierzawa95,Baeriswyl00}, and quantum Monte Carlo is
also applicable~\cite{Baeriswyl09,Eichenberger09}.  In the limit of large
interaction it is known that the BWF corresponds to bound excitons and is
therefore insulating~\cite{Baeriswyl00}.

The GWF is often treated via a combinatorical approximation also due to
Gutzwiller~\cite{Gutzwiller65,Brinkman70,Vollhardt84,Fazekas99,Edegger07}
(GA).  The GA predicts a metal-insulator transition (Brinkman-Rice
transition)~\cite{Gutzwiller65,Brinkman70,Vollhardt84,Fazekas99,Edegger07}
and is exact in infinite dimensions~\cite{Metzner89}.  The relation of the GA
to the exact GWF has also been studied~\cite{Kurzyk07,Hetenyi09a}.  In 
recent work the author and co-workers suggested that the GA consists of using
a simplified form for the spin correlations in the non-interacting reference
wave function~\cite{Hetenyi09a}.  Similar approaches, in which the exchange
interaction is implemented in an effective way, have also been used in
continuous systems to obtain approximations for
correlation.~\cite{Lado67,Stevens73,Hetenyi04,Hetenyi08} Extensions of the GA
include the time-dependent case~\cite{Seibold01}, implementation for the
multi-band models~\cite{Bunemann07}, ensembles with varying particle number
(BCS wavefunction)~\cite{Edegger05}, and the calculation of matrix elements
between ground and excited states~\cite{Fukushima05}.  The GA has also
recently been applied to fermions in optical
lattices~\cite{Heiselberg09,Wang08,Wang09}, combination with DFT
(LDA)~\cite{Deng08,Deng09,Wang10}, and
RPA~\cite{Seibold04,DiCiolo09,Markiewicz10a,Markiewicz10b}.  An improved
version of the GA was also recently proposed~\cite{Jedrak10}. In the context
of high tempereature superconductivity variants of the approximate solution
have been applied~\cite{Zhang88,Edegger06,Gros06} in the resonating valence
bond (RVB) method~\cite{Anderson73,Fazekas74,Anderson04}, which is based on a
completely projected Gutzwiller wavefunction.

For the BWF or its extensions (BGWF or GBWF) there has not been an
approximate solution of a similar type.  The aim of this work is to develop
such a scheme for the BWF and the BGWF.  The assumptions in the GA for the
spatial distribution are applied here in momentum space, an approximation for
the ${\bf k}$-space analog of the exchange hole, defined as
$\langle\Psi_G(\infty)|n_{{\bf k}\sigma}n_{{\bf
    k'}\sigma'}|\Psi_G(\infty)\rangle,$ (where $\Psi_G(\infty)$ denotes the
completely projected Gutzwiller wavefunction) is invoked.  The phase diagram
of the one-dimensional Hubbard model is calculated.  The phase transions are
characterized by a decrease of the Fermi step.  At half-filling it is a
metal-insulator transition (the Fermi step disappears), from an uncorrelated
metal to an insulator with finite double occupations.  Away from half-filling
the transition is between an uncorrelated metal (fully localized in momentum
space) to a correlated metallic state.  The correlated metallic state has a
finite Fermi step smaller than the Hartree-Fock solution.  The double
occupation tends to the value of the completely projected GWF, but double
occupations due to exciton binding are present at finite interaction.

Excitons in metals are rare due to screening by free charge carriers.
Recently evidence~\cite{Spataru04,Wang07} was found for the presence of bound
excitons in single-walled carbon nanotubes.  A carbon nanotube can be seen as
a system of low dimensionality, hence screening can be expected to be
significantly reduced, and the effects of correlations are more pronounced .
In these systems the experimental absorption lineshape can not be reproduced
by a tight-binding model alone, many electron effects are included via {\it
  GW}-type approaches~\cite{Spataru04}.  The variational ansatz presented
here incorporates bound excitons via an approximate variational theory.

This paper is organized as follows.  In the following section the method is
presented.  In particular the GA in its original form is used as a starting
point to construct a similar approximation for the BWF and the BGWF.  In
section \ref{sec:results} the results are presented.  Subsequently
conclusions are drawn.

\section{Method}
\label{sec:method}

\subsection{The Hubbard Hamiltonian and variational wavefunctions}
\label{ssec:HMGWF}

In this study the Hubbard
Hamiltonian~\cite{Gutzwiller63,Gutzwiller65,Hubbard63,Kanamori63} for
spin-unpolarized systems at various fillings will be investigated.  The
Hamiltonian in one dimension can be written
\begin{equation}
  H = -t \sum_{\langle i,j\rangle\sigma}^N c^\dagger_{i\sigma}c_{j\sigma} + U \sum_{i=1}^N n_{i\uparrow}n_{i\downarrow}.
\end{equation}
We will assume a system with $L$ lattice sites and with $N_\uparrow$ and
$N_{\downarrow}$ particles with spins up and down respectively.  The idea of
the BWF~\cite{Baeriswyl86,Baeriswyl00} is to act with a kinetic energy
projection operator on the completely projected GWF.  The GWF is obtained by
projecting out double occupations from a Fermi sea,
\begin{equation}
|\Psi_G(\gamma) \rangle = \mbox{exp}\left(-\gamma \sum_i
n_{i\uparrow}n_{i\downarrow}\right)|FS\rangle,
\label{eqn:Psi_G}
\end{equation}
where $|FS\rangle$ indicates a Fermi sea of non-interacting fermions.

The BWF can be defined using Eq. (\ref{eqn:Psi_G}) as
\begin{equation}
|\Psi_B(\alpha) \rangle = \mbox{exp}\left(-\alpha \left(\sum_{\langle
    i,j\rangle\sigma}^N c^\dagger_{i\sigma}c_{j\sigma}-\sum_{i\sigma}\mu_\sigma
n_{i\sigma}\right) \right)|\Psi_G(\gamma\rightarrow\infty)\rangle.
\label{eqn:Psi_B}
\end{equation}
Diagonalizing the hopping operator one can also write
\begin{equation}
|\Psi_B(\alpha) \rangle = \mbox{exp}\left(-\alpha \sum_{{\bf k}\sigma}^N
  (\epsilon({\bf k})-\mu_\sigma)n_{{\bf k}\sigma}\right)|\Psi_G(\gamma\rightarrow\infty)\rangle,
\label{eqn:Psi_B2}
\end{equation}
with $\epsilon({\bf k}) = -2t\mbox{cos}({\bf k})$ and $\mu_\sigma$ being the
chemical potential.  The completely projected GWF at half-filling contains no
double occupations.  For finite $\alpha$, however, double occupations arise
as a result of the binding of neighboring up-spin and down-spin particles and
their second order hopping processes, as shown by
Baeriswyl~\cite{Baeriswyl00}.  In particular Baeriswyl has
shown~\cite{Baeriswyl00} that the polarization fluctuations at half-filling
have the form
\begin{equation}
\langle X^2 \rangle = 2 \sum_{\langle i,j \rangle} \frac{t^2}{U^2}\left(
  \frac{1}{4} - \langle \Psi_G(\infty)|{\bf S}_i\cdot{\bf S}_j| \Psi_G(\infty)\rangle\right),
\end{equation}
where $X$ denotes the total position operator ${\bf S}_i$ denotes the spin
vector of site $i$.  This expression corresponds to bound pairs of double
occupations and holes or dipoles with random orientations.  Double
occupations arise as a result of second-order hopping processes.  It is also
interesting to note that the BWF is closely related to the
RVB~\cite{Anderson73,Fazekas74,Anderson04}.  In the RVB a completely
projected GWF is acted on by a unitary operator whose exponent consists of a
sum of selective hopping processes (increase or decrease of double
occupations).  The approximate solution of this method leads to solving a
spin-$\frac{1}{2}$ Heisenberg Hamiltonian, which is also true for the
BWF~\cite{Baeriswyl00}.  In the BWF, however, all hoppings are included in
the projection, hence away from half-filling the charge carriers can be
expected to be more mobile.

The two projections detailed above can also be applied in sequence.  Two
other variational wavefunctions can be obtained by
\begin{equation}
|\Psi_{BG}(\alpha,\gamma) \rangle = \mbox{exp}\left(-\alpha \left(\sum_{\langle
    i,j\rangle\sigma}^N c^\dagger_{i\sigma}c_{j\sigma}-\sum_{i\sigma}\mu_\sigma
n_{i\sigma}\right) \right)|\Psi_G(\gamma)\rangle,
\label{eqn:Psi_BG}
\end{equation}
with $\gamma$ finite, and 
\begin{equation}
|\Psi_{GB}(\alpha,\gamma) \rangle = \mbox{exp}\left(-\gamma \sum_i
n_{i\uparrow}n_{i\downarrow}\right)|\Psi_B(\alpha)\rangle.
\label{eqn:Psi_GB}
\end{equation}
Eq. (\ref{eqn:Psi_BG})(Eq. (\ref{eqn:Psi_GB})) is known as the
Baeriswyl-Gutzwiller~\cite{Otsuka92,Dzierzawa95}(Gutzwiller-Baeriswyl)
wavefunction.  Below, in addition to the BWF, an approximation scheme is also
developed for the Baeriswyl-Gutzwiller wavefunction (BGWF).

\subsection{The Gutzwiller approximation}
\label{ssec:GA}

In the following the essential features of the GA will be given, for details
see Refs. \onlinecite{Gutzwiller65,Brinkman70,Vollhardt84,Fazekas99}.  The GA
was developed to simplify the sum over determinants that arise when
expectation values are evaluated over $\Psi_G$.  The approximations are based
on the $U=0$($\gamma=0$) solution.  In the position representation the
normalization of the GWF can be written
\begin{equation}
  \langle \Psi_G | \Psi_G \rangle = L^{-\left( N_\uparrow+N_\downarrow \right) }
  \sum_I
  |{\mathfrak D}[{\bf k;g}_I]|^2 
  |{\mathfrak D}[{\bf l;h}_I]|^2\mbox{exp}[-2\gamma D({\bf g}_I,{\bf h}_I)], 
\end{equation}
where the sum is over all configurations of coordinates, ${\bf g}_I$ and
${\bf h}_I$ denote the coordinates corresponding to configuration $I$,
$D({\bf g}_I,{\bf h}_I)$ denotes the number of double occupations for the
particular configuration, and ${\mathfrak D}[{\bf k;g}_I]$ denotes the
determinant formed of the plane-waves with wavevectors ${\bf k}$ at positions
${\bf g}_I$.  Due to the determinants only those configurations contribute
which include up to one particle of a particular spin at each site.  One can
define the unnormalized probability distribution in position space,
\begin{equation}
  P_{GWF}({\bf g},{\bf h}) = 
|{\mathfrak D}[{\bf k;g}]|^2|{\mathfrak D}[{\bf l;h}]|^2
\mbox{exp}[-2\gamma D({\bf g},{\bf h})].
\label{eqn:P_GWF}
\end{equation}
Using Eq. (\ref{eqn:P_GWF}) one can write relevant expectation values.  For
example the average double occupation can be written
\begin{equation}
\label{eqn:D_GWF}
  \left\langle \sum_i n_{i\uparrow}n_{i\downarrow} \right\rangle =
  \frac{\sum_{I} P_{GWF}({\bf g}_I,{\bf h}_I) D({\bf g}_I,{\bf h}_I)}
  {\sum_{I} P_{GWF}({\bf g}_I,{\bf h}_I)}.
\end{equation}

To arrive at the GA one replaces the square of the determinants with their
averages in the Fermi sea.  Considering only the up-spin channel one can
write the normalization of the Fermi sea as
\begin{equation}
  _{\uparrow}\langle FS | FS \rangle_{\uparrow} =
  L^{-N_\uparrow} \sum_I |{\mathfrak D}[{\bf k;g}_I]|^2 = 1,
\label{eqn:norm_up}
\end{equation}
since the wavefunctions that enter are normalized planewaves themselves.  As
the sum in Eq. (\ref{eqn:norm_up}) is over all configurations of up-spin
particles on the lattice, such that at most one particle occupies a
particular site we can approximate each term by its average as
\begin{equation}
  |{\mathfrak D}[{\bf k;g}_I]|^2 \approx \langle |{\mathfrak D}[{\bf k;g}_I]|^2\rangle = \frac{L^{N_\uparrow}}{C^L_{N_\uparrow}},
\label{eqn:approx_norm_sl}
\end{equation}
where $C^L_{N_\uparrow}$ denotes the number of ways $N_\uparrow$ particles
can be placed on $L$ lattice sites.  The down-spin particles can be handled
similarly.  Substituting Eq. (\ref{eqn:approx_norm_sl}) one can write the
unnormalized probability distribution in real space as
\begin{equation}
  P_{GA}({\bf g}{\bf h}) = 
\mbox{exp}[-2\gamma D({\bf g},{\bf h})].
\label{eqn:P_GA}
\end{equation}
The average number of double occupations in terms of $P_{GA}$ can be written
\begin{equation}
\label{eqn:D_GA}
  \left\langle \sum_i n_{i\uparrow}n_{i\downarrow} \right\rangle =
  \frac{\sum_I P_{GA}({\bf g}_I,{\bf h}_I) D({\bf g}_I,{\bf h}_I)}
  {\sum_I P_{GA}({\bf g}_I,{\bf h}_I)},
\end{equation}
but here, contrary to Eq. (\ref{eqn:D_GWF}) a constraint has to be introduced
over the summations.  Only those configurations are summed over, which have
zero or one particle of a particular spin at each lattice site.

The approximation for the kinetic energy involves averaging a product of
unequal
determinants~\cite{Gutzwiller65,Brinkman70,Vollhardt84,Fazekas99,Yokoyama87,Hetenyi09a}
since the hopping is not diagonal in the coordinate representation.  Similar
to Eq. (\ref{eqn:approx_norm_sl}) this is done by evaluating the hopping
energy for the Fermi sea,
\begin{equation}
\label{eqn:t_up1}
T =_{\uparrow}\langle FS |c^{\dagger}_{i\uparrow} c_{j\uparrow}|
FS \rangle_{\uparrow} 
= L^{-N_\uparrow}\sum_I\hspace{.001cm}'\hspace{.1cm}
{\mathfrak D}^*[{\bf
  k;g'}_I]{\mathfrak D}[{\bf k;g}_I], 
\end{equation}
where ${\bf g}_I$ and ${\bf g'}_I$ denote two configurations which differ
only in their occupations of site $i$ and $j$.  For ${\bf g}_I$(${\bf g'}_I$)
site $i$ is unoccupied(occupied) and site $j$ is occupied(unoccupied).  The
prime on the sum indicates the restriction that ${\bf g}_I$ are
configurations such that site $i$ is occupied and site $j$ unoccupied.  Of
configurations with a given pair of sites which have one occupied and one
unoccupied site there are $C^{L-2}_{N_\uparrow-1}$.  The product of
determinants can be approximated by the Fermi sea average, since the hopping
energy can be evaluated exactly, i.e.
\begin{equation}
\label{eqn:t_up2}
T = \frac{1}{L}{\sum_{\bf k}}^* \mbox{exp}[i {\bf k\cdot (R_i
  - R_j)})].
\end{equation}
where ${\bf R_i}$ and ${\bf R_j}$ denote the pair of lattice sites involved
in the hopping, and the asterisk indicates that the sum be performed over
occupied states only.  The approximation
\begin{equation}
{\mathfrak D}^*[{\bf  k;g'}_I]{\mathfrak D}[{\bf k;g}_I]
\approx
\langle {\mathfrak D}^*[{\bf  k;g'}_I]{\mathfrak D}[{\bf k;g}_I] \rangle
= T\frac{L^{N_\uparrow}}{C^{L-2}_{N_\uparrow-1}}
\end{equation}
can be introduced.  Using this approximation the average hopping of an
up-spin particle from site $j$ to site $i$ can be written
\begin{equation}
\label{eqn:hop_GA}
\frac{\langle \Psi |c^\dagger_{i\uparrow}c_{j\uparrow} |\Psi \rangle}
{\langle \Psi | \Psi \rangle}
\approx  T \frac{C^L_{N_\uparrow}}{C^{L-2}_{{N_\uparrow}-1}} 
\frac{
  \sum_I'
  P_{GA}({\bf g}_I,{\bf h}_I)\mbox{exp}[-\gamma \Delta D({\bf g'}_I,{\bf g}_I;{\bf h}_I)]
}
{
  \sum_I
  P_{GA}({\bf g}_I,{\bf h}_I)
}.
\end{equation}
In Eq. (\ref{eqn:hop_GA}) $\Delta D$ indicates the change in number of double
occupations due to the hopping.  

Notice that one could arrive at the approximate expressions in
Eqs. (\ref{eqn:P_GA}) and (\ref{eqn:hop_GA}) via different
reasoning~\cite{Hetenyi09a}.  One can define a probability distribution over
configurations with up to one particle of a given spin at each site and weigh
each configuration with the weighing factor $P_{GA}({\bf g}_I,{\bf
  h}_I)=\mbox{exp}(-2\gamma D({\bf g}_I,{\bf h}_I))$.  One can define an
estimator for the hopping energy of the form $ \tilde{T}\mbox{exp}(-\gamma
\Delta D({\bf g'}_I,{\bf g}_I;{\bf h}_I))$, considering that the hopping
operator connects states with different number of double occupations.  The
scaling factor $\tilde{T}$ can be obtained by requiring that the hopping
energy at $U=0$ ($\gamma=0$) corresponds to the kinetic energy of the
non-interacting system (Eq. (\ref{eqn:t_up2})).  Below the approximation
scheme for the BWF will follow these steps.

The Gutzwiller approximation gives rise to a very simple form for the momentum
distribution 
\begin{equation}
  \langle n_{{\bf k}\sigma} \rangle_\gamma = n_\sigma(1-q(\gamma)) + q(\gamma) \Theta(\mu_\sigma-\epsilon_{\bf k}),
\label{eqn:nks1}
\end{equation}
where $\mu_\sigma$ and $\Theta(x)$ denotes the chemical potential and the
Haviside step function respectively, and 
\begin{equation}
q(\gamma) = \frac{C^L_{N_\uparrow}}{C^{L-2}_{{N_\uparrow}-1}} 
  \frac{
\sum_I
P_{GA}({\bf g}_I,{\bf h}_I)
\mbox{exp}[-\gamma \Delta D({\bf g'}_I,{\bf g}_I;{\bf h}_I)]
}
{
\sum_I
P_{GA}({\bf g}_I,{\bf h}_I)
}.
\label{eqn:nks2}
\end{equation}
From Eq. (\ref{eqn:nks1}) one sees that the momentum distribution at any
filling is a constant function with a discontinuity at the value of the
chemical potential.  For half-filling in the limit $\gamma \rightarrow
\infty$ the distribution becomes $\frac{1}{2}$ for any ${\bf k}$, and
$q(\gamma)$ can be simplified~\cite{Fazekas99} to
\begin{equation}
q(\gamma) = \frac{4 \mbox{exp}(-\gamma)}{(1+\mbox{exp}(-\gamma))^2}.
\label{eqn:nks3}
\end{equation}

\subsection{Application to the Baeriswyl wavefunction}
\label{ssec:GA-B}

The BWF consists of a projection of the fully projected GWF.  While the
normalization for the GWF can be easily written, since the $U=0$ solution is
known, this is more difficult for the BWF where the $U=\infty$ solution is
needed.  In general one can write the normalization as
\begin{equation}
  \langle \Psi_B | \Psi_B \rangle = \sum_I
\chi({\bf k}_I,{\bf l}_I)
  \mbox{exp}\left[-2\alpha \sum_I \{\epsilon({\bf k}_I)+\epsilon({\bf l}_I) \}  \right],
\label{eqn:norm}
\end{equation}
where
\begin{equation}
\chi({\bf k}_I,{\bf l}_I) =
\langle \Psi_G(\gamma\rightarrow\infty)|{\bf k}_I,{\bf l}_I \rangle
\langle {\bf k}_I,{\bf l}_I|\Psi_G(\gamma\rightarrow\infty) \rangle.
\end{equation}
$\chi({\bf k}_I,{\bf l}_I)$ denotes a probability distribution for a
particular set of vectors ${\bf k}_I$ and ${\bf l}_I$ which guarantees that
at $\alpha=0$ the $\gamma=\infty$ momentum distribution
(Eq. (\ref{eqn:nks1})) is recovered.  To account for this distribution one
introduces a piecewise constant potential discontinuous at $\mu_\sigma$ in
momentum space, so that the distribution in Eq. (\ref{eqn:nks1}) is
reproduced.  As in the GA the summation in Eq. (\ref{eqn:norm}) is such that
no two particles of the same spin can occupy the same site in momentum space
(to account for the Pauli principle), but the distribution is otherwise
uncorrelated.  The kinetic energy is obtained the usual way
\begin{equation}
  \left\langle \sum_{{\bf k}}\epsilon({\bf k})\tilde{n}_{{\bf k},\sigma}
  \right \rangle
  = 
\frac{ 
\sum_I
P_{GA-B}({\bf k}_I,{\bf l}_I)
\{\sum_I
(\epsilon({\bf k}_I)+\epsilon({\bf l}_I))\}
}
{
\sum_I
P_{GA-B}({\bf k}_I,{\bf l}_I)
},
\end{equation}
with $P_{GA-B}$ defined as
\begin{equation}
P_{GA-B}({\bf k}_I,{\bf l}_I)
= \chi({\bf k}_I,{\bf l}_I)
\mbox{exp}[-2\alpha \sum_{{\bf k}_I,{\bf l}_I}(\epsilon({\bf k}_I)+\epsilon({\bf l}_I))].
\end{equation}

In order to arrive at an approximation scheme for the interaction we first
write the number of double occupations in ${\bf k}$-space as
\begin{equation}
\sum_i n_{i\uparrow}n_{i\downarrow} = \frac{1}{L} 
\left( \sum_{\bf k,k'} \tilde{n}_{{\bf k}\uparrow}\tilde{n}_{{\bf
      k'}\downarrow} - 
\sum_{{\bf k k' q }\neq0} 
\tilde{c}_{{\bf k}\uparrow}^\dagger
\tilde{c}_{{\bf k'}\downarrow}^\dagger
\tilde{c}_{{\bf k+q}\uparrow}
\tilde{c}_{{\bf k'-q}\downarrow}
\right).
\label{eqn:Dk}
\end{equation}
The first term is simply $N_\uparrow N_\downarrow /L$.  The second term is a
correlated hopping of an up-spin and down-spin particle in momentum space.
We write
\begin{equation}
\label{eqn:D2}
  \left\langle\sum_{{\bf k k' q }\neq0} 
    \tilde{c}_{{\bf k}\uparrow}^\dagger
    \tilde{c}_{{\bf k'}\downarrow}^\dagger
    \tilde{c}_{{\bf k+q}\uparrow}
    \tilde{c}_{{\bf k'-q}\downarrow}
  \right \rangle
  =
  \tilde{U}\frac{ \sum_I''
    P_{GA-B}({\bf k}_I,{\bf l}_I)
    \sum_{{\bf k_I,k'_I,q}\neq0}\mbox{exp}\{-\alpha(\epsilon({\bf k}_I)+\epsilon({\bf k'}_I)-
    \epsilon({\bf k}_I+{\bf q})-\epsilon({\bf k'}_I-{\bf q}))\}
  }
  {\sum_I
P_{GA-B}({\bf k}_I,{\bf l}_I)
}
\end{equation}
where the double prime indicates that for a particular set ${\bf k_I,k'_I,
  q}\neq0$ the only configurations which enter are ones with ${\bf k'}_I-{\bf
  q}$ and ${\bf k'}_I+{\bf q}$ unoccupied and ${\bf k}_I$ and ${\bf k'}_I$
occupied.  The energy differences in the exponent account for the correlated
hopping in momentum space.  In the original GA applied to the GWF, it is the
hopping energy which behaves in a similiar way: there the hopping causes a
change in the number of double
occupations~\cite{Gutzwiller65,Brinkman70,Vollhardt84,Fazekas99}.
$\tilde{U}$ is fixed by requiring that the known number of double occupations
is reproduced at $U=\infty$ ($\alpha=0$), in other words the ${\bf q}=0$ term
is cancelled by the ${\bf q}\neq0$ one.  Note that in the GA the kinetic
energy is multiplied by a scaling factor which is fixed by requiring that the
noninteracting kinetic energy is reproduced~\cite{Hetenyi09a}.

\begin{figure}[htp]
\vspace{1cm}
\psfig{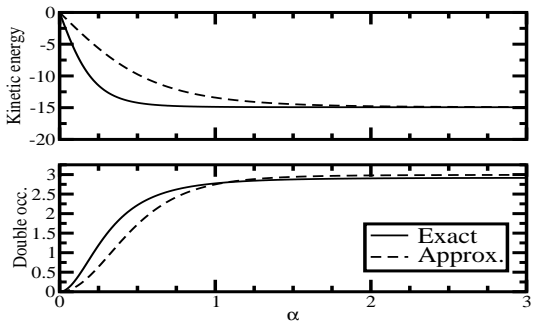}
\vspace{1cm}
\caption{ Kinetic and potential energies as a function of the variational
  parameter at half-filling. }
\label{fig:nrg_comp1}
\end{figure}

\begin{figure}[htp]
\vspace{1cm}
\psfig{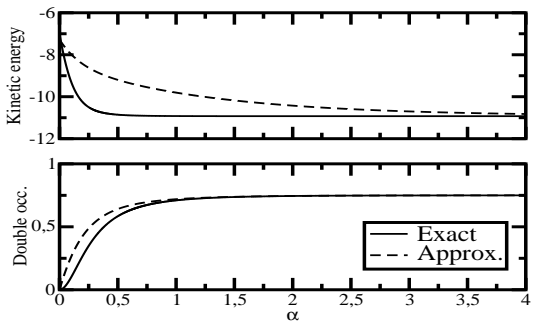}
\vspace{1cm}
\caption{ Kinetic and potential energies as a function of the variational
  parameter at quarter-filling. }
\label{fig:quarter_nrg_comp1}
\end{figure}

In the limit $\alpha\rightarrow\infty$ the Hartree-Fock non-interacting
ground state is obtained.  In Eq. (\ref{eqn:norm}) the distribution in this
limit includes only the ${\bf k}$ vectors corresponding to the lowest
eigenvalues.  Since the distribution corresponds to the finite temperature
one with inverse temperature equal to $2\alpha$, it also follows that at half
filling the discontinuity characterizing the Fermi surface of metals is only
present when $\alpha\rightarrow\infty$.  For the double occupations, only the
${\bf q}=0$ term survives when $\alpha\rightarrow\infty$.  In the limiting
cases $U=0$ and $U\rightarrow\infty$ the energies are correct within the
present scheme, as is the case for the standard GA applied to the GWF.
Generalization to systems away from half-filling is also straightforward,
since the distribution for the $\alpha \rightarrow 0$ limit ($\chi({\bf
  k},{\bf l}$)) can be chosen accordingly.

\subsection{Application to the Baeriswyl-Gutzwiller wavefunction}
\label{ssec:GA-BG}

It is also possible to generalize the above scheme to the combined projection
based BGWF~\cite{Otsuka92,Dzierzawa95} defined in Eq. (\ref{eqn:Psi_BG} ).
This generalization is enabled by the fact that the completely projected GWF
on which the BWF is based (see Eqs. (\ref{eqn:Psi_G}) and (\ref{eqn:Psi_B}))
enters into the approximation scheme detailed in the previous subsection via
the momentum density, which for the GA has a known form
(Eq. (\ref{eqn:nks1})).  Hence generalizing the approximation scheme to the
BGWF proceeds exactly as described above, only that the momentum density is
the one corresponding to the value of $\gamma$ (Eq. (\ref{eqn:nks1})).

\subsection{Implementation}

In this work our implementation of the above approximations is similar to
that described in Chapter 9 of Ref. \onlinecite{Fazekas99}.  Assuming that
the distribution is fully uncorrelated in ${\bf k}$-space one can write the
normalization as
\begin{equation}
  \langle \Psi_B | \Psi_B \rangle \approx \prod_{{\bf k}\sigma} [ 1 + \langle
  n_{{\bf
    k}\sigma}\rangle_\gamma\{\mbox{exp}(-2 \alpha \epsilon({\bf k})) - 1\}].
\label{eqn:norm_imp}
\end{equation}
$\langle n_{{\bf k}\sigma}\rangle_\gamma$ is the momentum distribution of the
Gutzwiller wavefunction (Eq. (\ref{eqn:nks1})).  The kinetic energy is then
obtained via
\begin{equation}
  \sum_{{\bf k}\sigma}\epsilon({\bf k})\langle n_{{\bf k}\sigma} \rangle = - \frac{1}{2}\frac{\partial}{\partial\alpha} \mbox{ln} \langle
  \Psi_B | \Psi_B \rangle = \sum_{{\bf k}\sigma} 
\frac
{\mbox{exp}(-2 \alpha \epsilon({\bf k})) \epsilon({\bf k})  \langle n_{{\bf
      k}\sigma} \rangle_\gamma } 
{1 + \{\mbox{exp}(-2 \alpha \epsilon({\bf k})) - 1\}  \langle n_{{\bf
      k}\sigma}  \rangle_\gamma}.
\end{equation}
The resulting interaction energy 
\begin{eqnarray}
\label{eqn:docc}
  U \sum_i\langle n_{i\uparrow}n_{i\downarrow} \rangle &\approx &
  U \frac{N_\uparrow N_\downarrow}{L} \\
 & &
  + U \tilde{U}\sum_{{\bf k,k',q}\neq0} 
\frac{
\mbox{exp}\{-\alpha(\epsilon({\bf k})+\epsilon({\bf k'})-
    \epsilon({\bf k}+{\bf q})-\epsilon({\bf k'}-{\bf q}) \} 
(1 - \langle n_{{\bf k}+{\bf q}  \uparrow}\rangle_\gamma)(1-\langle n_{{\bf
    k'}-{\bf q}\downarrow} \rangle_\gamma)
       \langle n_{{\bf k}\uparrow} \rangle_\gamma \langle n_{{\bf
           k'}\downarrow} \rangle_\gamma
}
{ C({\bf k},{\bf k'},{\bf q}) } \nonumber
\end{eqnarray}
where
\begin{eqnarray}
C({\bf k},{\bf k'},{\bf q}) &=&
(1 + \langle n_{{\bf k}\sigma} \rangle_\gamma\{\mbox{exp}(-2 \alpha \epsilon({\bf k})) - 1\})
(1 + \langle n_{{\bf k'}\sigma}\rangle_\gamma\{\mbox{exp}(-2 \alpha \epsilon({\bf k'})) - 1\}) \\
& &
(1 + \langle n_{{\bf k}+{\bf q}\sigma} \rangle_\gamma \{\mbox{exp}(-2 \alpha \epsilon({\bf k}+{\bf q})) - 1\})
(1 + \langle n_{{\bf k'}+{\bf q}\sigma}\rangle_\gamma \{\mbox{exp}(-2 \alpha \epsilon({\bf k'}+{\bf
  q})) - 1\}). \nonumber
\end{eqnarray}
Note that the occupation factors in Eq. (\ref{eqn:docc}) are such that states
${\bf k}$ and ${\bf k'}$ are occupied, and ${\bf k}+{\bf q}$ and ${\bf
  k'}-{\bf q}$ are unoccupied, which coincide with the correlated hoppings in
${\bf k}$-space corresponding to the double occupation operator.

\section{Results}

\label{sec:results}

\begin{figure}[htp]
\vspace{1cm}
\psfig{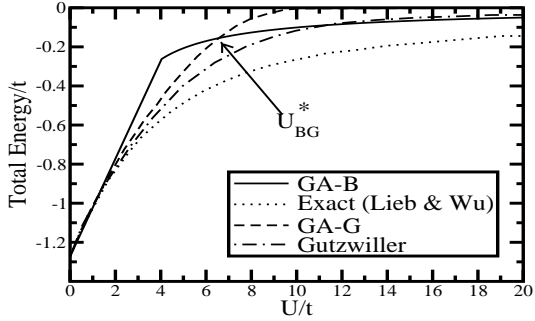}
\vspace{1cm}
\caption{Comparison of the energy of the GA-B scheme to well known results:
  GA-G (Gutzwiller approximation applied to the GWF), exact Gutzwiller, and
  exact results. $U^*_{BG}$ indicates the transition point for the
  Baeriswyl-Gutzwiller wavefunction.}
\label{fig:nrg_comp2}
\end{figure}

In Figs. \ref{fig:nrg_comp1} and \ref{fig:quarter_nrg_comp1} the kinetic and
interaction energies are shown for a twelve site system comparing the results
of an exact calculation to the outcome of the approach presented here for
half and quarter fillings.  The kinetic energy shows strong disagreement for
intermediate values of the variational parameter, presumably due to the fact
that this approach does not take into account momentum space correlations.
The double occupations are in good agreement between the two calculations in
both cases.  A similar degree of agreement is found at quarter filling.
Further testing of the method can be seen in Fig. \ref{fig:nrg_comp2} where
the energy at half-filling is compared to the exact result~\cite{Lieb68}, the
exact Gutzwiller result~\cite{Metzner87,Metzner89} and the Gutzwiller
approximation applied to the GWF at half-filling.  The discontinuity in the
GA-B results indicates a first-order metal-insulator transition at
$U_c\approx4.04$, where the $U<U_c$ phase corresponds to
$\alpha\rightarrow\infty$, the Hartree-Fock solution.

\begin{figure}[htp]
\vspace{1cm}
\psfig{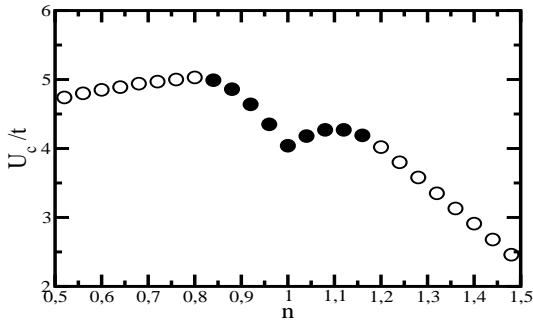}
\vspace{1cm}
\caption{Critical interaction strength as a function of particle
  density. Closed circles indicate systems in which the large interaction
  phase contains bound excitons. }
\label{fig:pd}
\end{figure}

\begin{figure}[htp]
\vspace{1cm}
\psfig{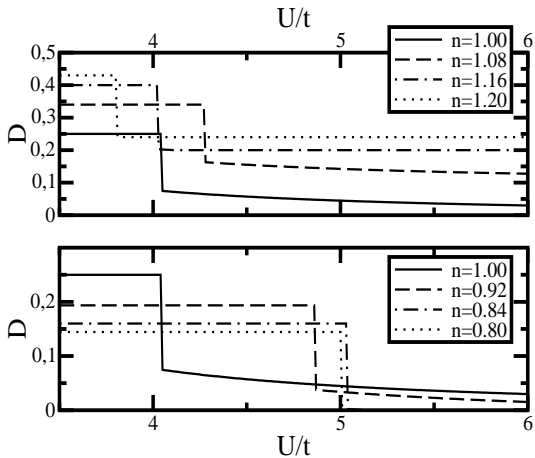}
\vspace{1cm}
\caption{Double occupation as a function of interaction strength for various
  fillings. }
\label{fig:d}
\end{figure}

In Fig. \ref{fig:pd} the phase diagram is presented calculated using $200$
sites.  As the density decreases from half-filling, the critical interaction
strength increases until it reaches a maximum.  Similar behaviour is found
when the density is increased from half-filling.  In Fig. \ref{fig:d} the
fraction of doubly occupied sites are shown as a function of the interaction
strength at different fillings.  For the half-filling case (shown in both
panels) double occupations starts at one quarter (Hartree-Fock value), and
then decreases abruptly at $U_c$.  Subsequently it decays to zero with
increasing $U$.  This transition "mirrors" the
Brinkman-Rice~\cite{Brinkman70} transition.  There, while approaching the
critical interaction from the metallic side, the number of double occupations
decreases.  The insulator of the Brinkman-Rice transition is the simplest
possible insulator, with no double occupations.  In the GA-B the double
occupations increase when approaching the critical interaction from the
insulating side, and the metallic side corresponds to the simplest metal; the
non-interacting Hartree-Fock ground state.

Fig. \ref{fig:d} also shows how the fraction of double occupied sites vary
for different fillings.  For $n<1$, close to half-filling the double
occupations show the same pattern as for half-filling, until at $n\approx0.8$
the state corresponding to large interaction strength no longer contains
doubly occupied sites.  Above half-filling there is a minimum fraction of
doubly occupied sites for each system, but close to half-filling we observe a
slow tending to the large $U$ limiting value (for example $n=1.08$).  The
fraction of doubly occupied sites above the limiting value are due to bound
excitons, as they would not arise were it not for the Baeriswyl projector.
Such bound excitons were only found in the range $n=0.8 \leq n \leq 1.20$.
\begin{figure}[htp]
\vspace{1cm}
\psfig{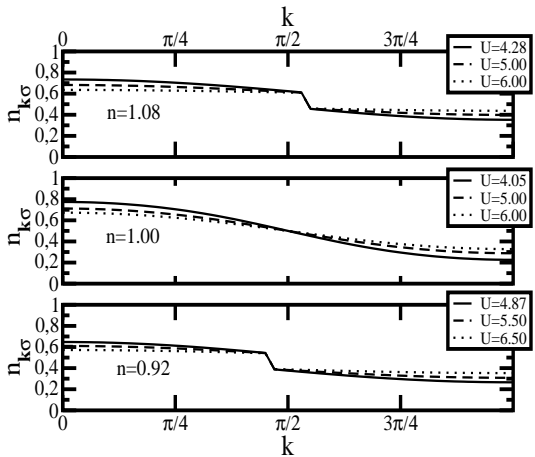}
\vspace{1cm}
\caption{Density as a function of $k$-vector at different fillings for
  various values of the interaction strength. }
\label{fig:nk}
\end{figure}

In Fig. \ref{fig:nk} the momentum density $n_{{\bf k}\sigma}$ is shown for
different fillings as a function of $U$ in the phase corresponding to large
$U$ in each case.  The phases found at small $U$ have a Fermi step of $1$ at
$\pi/2$.  The Fermi step closes entirely for the system at half-filling, but
remains finite for both systems away from half-filling.  Thus away from
half-filling, we find a a first order phase transition from an uncorrelated
metallic phase to a metallic phase which contains bound excitons.

\begin{figure}[htp]
\vspace{1cm}
\psfig{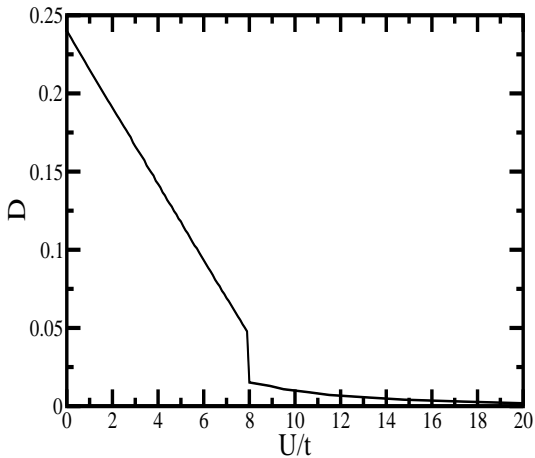}
\vspace{1cm}
\caption{Double occupation as a function of interaction strength for the
  Baeriswyl-Gutzwiller wavefunction for filling $n=0.98$.}
\label{fig:dbg}
\end{figure}

For the Baeriswyl-Gutzwiller projection the approximate scheme presented here
results in a minimum energy corresponding either to the Gutzwiller or the
Baeriswyl wavefunction.  For half-filling the transition occurs between a
correlated metal and a correlated insulator.  The interaction strength at
which the transition occurs is given by the crossing point of the energy
curves GA-G and GA-B, and is indicated in Fig. \ref{fig:nrg_comp2}
($U^*_{BG}\approx 6.6$).  Away from half-filling the interaction strength at
which the transition occurs increases, and for $n=0.96$ we find no transition
in the range $0\leq U\leq 20$: the ground state of the system is a partially
projected Gutzwiller function( no bound excitons).  For $n=0.98$ a first
order phase transition is found.  The transition occurs at
$U^*_{BG}\approx7.9$.  For smaller values of the interaction the wavefunction
is a partially projected Gutzwiller function, the Baeriswyl projection
parameter ($\alpha$) is always zero, only the Gutzwiller parameter ($\gamma$)
varies: the system is a correlated metal without bound excitons.  For larger
values of $U$ the parameter $\alpha$ is finite and approaches zero as
$U\rightarrow\infty$.  The Gutzwiller parameter is such that all double
occupations are projected out for $\alpha=0$, and is constant in this range
of $U$.  In Fig. (\ref{fig:dbg}) the double occupation is shown as a function
of the interaction strength.  The double occupation decreases linearly for
the correlated metal described by the Gutzwiller approximation, and is
discontinuous at the phase transition.  For the large interaction the double
occupation decays to zero.  In the regime where $\alpha$ and the double
occupation is finite the double occupations can be attributed to bound
excitons.  Hence away from half-filling a metal-metal transition is found
between two correlated metallic states, distinguished by the absence or
presence of bound excitons.

\section{Conclusion}

\label{sec:summary}

In summary, an approximate scheme was presented to solve the Baeriswyl and
Baeriswyl-Gutzwiller variational wavefunctions.  The approach presented here
is simple and easy to apply in finite dimensional systems and large system
sizes are tractable.  The scheme is similar in spirit to the well-known
Gutzwiller approximation, in which the starting point is the Fermi sea, and
the Pauli principle is implemented by requiring that no two particles of the
same spin can be on the same site in real space, but no other spin
correlation effects are included.  In the approach described herein two
particles of the same spin cannot occupy the same site in ${\bf k}$-space,
hence an approximate treatment of the ${\bf k}$-space analog of the exchange
hole is developed.

At half-filling a metal-insulator transition is found, where the metallic
phase ($U<U_c$) corresponds to the Hartree-Fock solution, the insulating
phase is one with finite double occupations corresponding to bound excitons.
This transition can be viewed as the "inverse" of the Brinkman-Rice
transition.  Close to but away from half filling, the $U>U_c$ phase displays
a finite Fermi step (metallic), as well as double occupations originating
from bound excitons.  As the filling is increased or decreased from
half-filling bound excitons are supressed.

For the Baeriswyl-Gutzwiller wavefunction it was found that the optimal
solution is always either the Baeriswyl or the Gutzwiller wavefunction in
this approximate scheme.  The phase transitions shift to larger values of the
interaction strength.  At half-filling a metal-insulator transition occurs
between a correlated metal (with double occupations suppressed) and a
correlated insulator (double occupations arising from bound excitons).  Away
from, but still close to, half-filling a transition was found between two
metallic phases, the correlated metallic state arising from the Gutzwiller
approximation for small interaction, and one containing double occupations
arising from exciton binding for large interaction.

\begin{acknowledgements}
Part of this work was performed at the Institut f\"ur Theoretische Physik at
TU-Graz under FWF (F\"orderung der wissenschaftlichen Forschung) grant number
P21240-N16.  Part of this work was performed under the HPC-EUROPA2 project
(project number 228398).
\end{acknowledgements}


\end{document}